\documentclass[prl,twocolumn,superscriptaddress,altaffillsymbol]{revtex4-1}
\usepackage{amssymb}
\usepackage{amsmath}
\usepackage{commath}
\usepackage{graphicx}
\usepackage{verbatim}
\usepackage{epsfig}
\usepackage{epstopdf}
\usepackage{color}
\usepackage{siunitx}
\usepackage{array}
\usepackage{upgreek}
\newcolumntype{P}[1]{>{\centering\arraybackslash}p{#1}}
\newcolumntype{M}[1]{>{\centering\arraybackslash}m{#1}}

\DeclareSIUnit\gauss{G}

\begin{document}
	
\title{Probing magnon dynamics and interactions in a ferromagnetic spin-1 chain}

\author{Prashant Chauhan}
\affiliation{The Institute for Quantum Matter, Department of Physics and Astronomy, The Johns Hopkins University, Baltimore, Maryland 21218, USA}

\author{Fahad Mahmood}
\email{fahad@jhu.edu}
\affiliation{The Institute for Quantum Matter, Department of Physics and Astronomy, The Johns Hopkins University, Baltimore, Maryland 21218, USA}

\author{Hitesh J. Changlani}
\email{hchanglani@fsu.edu}
\affiliation{Department of Physics, Florida State University, Tallahassee, Florida 32306, USA}
\affiliation{National High Magnetic Field Laboratory, Tallahassee, Florida 32304, USA}
\affiliation{The Institute for Quantum Matter, Department of Physics and Astronomy, The Johns Hopkins University, Baltimore, Maryland 21218, USA}

\author{S.~M.~Koohpayeh}
\affiliation{The Institute for Quantum Matter, Department of Physics and Astronomy, The Johns Hopkins University, Baltimore, Maryland 21218, USA}

\author{N.~P.~Armitage}
\affiliation{The Institute for Quantum Matter, Department of Physics and Astronomy, The Johns Hopkins University, Baltimore, Maryland 21218, USA}


\begin{abstract}
	NiNb$_{2}$O$_{6}$ is an almost ideal realization of a 1D spin-1 ferromagnetic Heisenberg chain compound with weak unidirectional anisotropy. Using time-domain THz spectroscopy, we measure the low-energy electrodynamic response of NiNb$_{2}$O$_{6}$ as a function of temperature and external magnetic field. At low temperatures, we find a magnon-like spin-excitation, which corresponds to the lowest energy excitation at $q\sim0$.  At higher temperatures, we unexpectedly observe a temperature-dependent renormalization of the spin-excitation energy, which has a strong dependence on field direction. Using theoretical arguments, exact diagonalizations and 
finite temperature dynamical Lanczos calculations, we construct a picture of magnon-magnon interactions that naturally explains the observed renormalization. This unique scenario is a consequence of the spin-1 nature and has no analog in the more widely studied spin-1/2 systems. 
\end{abstract}

\maketitle
\setlength\belowcaptionskip{-3ex}

Since the early work of Ising (1925)~\cite{Ising_1925} and Bethe (1931)~\cite{Bethe_1931}, magnetism in 1D spin chains has been the subject of 
continuous theoretical~\cite{Bonner_1981, HALDANE_PRL_1983, WEINERT_1983,  affleck_1988, Affleck_JPCM_1989, Papaniculou_1987, Papaniculou_1997, Damle_1998,
Suzuki)PRB_2018, Sule_Changlani_2015, Richter_2019} and experimental interest~\cite{Steiner_1976, Katsumata_2000, Kimura_PRL_2007, Coldea_2010, 
Morris_PRL2014,Grenier_PRL_2015, Wang_PRB_2015, Wang_nat_2018, Faure_Natphy_2018}. Due to reduced dimensionality, magnetic order is susceptible to 
quantum fluctuations which can cause the system to exhibit interesting quantum effects~\cite{Mermin_Wagner_1966}. Examples include novel quantum phase transitions~\cite{sachdev_2011,Coldea_2010}, fractional excitations~\cite{McCoy_PRD_1978, FADDEEV_1981}, entanglement~\cite{Blanc_2018} and spin-charge separation~\cite{Kim_PRL_1996, Moreno_PRB_2013}. Moreover, the simplicity of 1D systems often makes the theoretical formulation tractable and allows a direct comparison with experiment.

Spin excitations in 1D chains have been studied for both ferromagnetic (FM) and antiferromagnetic (AFM) exchange interactions~\cite{HALDANE_PRL_1983, Katsumata_2000, Lissouck_2007,Morris_PRL2014}. For an isolated FM spin-1/2 chain with pure Ising interactions, the excitations ($ |m_z= 1/2\rangle\rightarrow |-1/2\rangle $) are domain walls. Each spin flip forms two domain walls (`kinks' or `spinons') carrying spin $s=1/2$~\cite{Rutkevich_2008, PFEUTY_1970}. These fractional excitations can be understood analytically and have been studied extensively in a variety of 1D spin-1/2 systems~\cite{Coldea_2010,Morris_PRL2014,Grenier_PRL_2015,Wang_PRB_2015}. 

The elementary excitation of a FM spin-1 chain is a magnon-like spin-flip $|1\rangle\rightarrow|0\rangle$. This excitation has a well-defined energy and momentum and is relatively easy to understand. However, magnon-magnon interactions are possible for a higher number of spin-flips leading to a renormalization of the spin excitation energies in ways that are quite distinct from the more commonly studied spin-1/2 chains. For example, as we will discuss below, in a spin-1 chain when two spin-flips (two magnons) come together, they can tunnel into other configurations like $|00\rangle\rightarrow |1$$-$$1\rangle $ and $ |$$-$$11\rangle$ to form a hybridized state which can alter the magnon spectrum. This process cannot occur in spin-1/2 chains. 

In general, the physics of such spin chains can be modeled with a nearest-neighbor exchange interaction $J$ and an in-plane anisotropy strength $D$. For spin-1, 
weakly anisotropic chains $D$$<$$J$ with AFM interactions, one obtains the `Haldane gap' which has been the subject of extensive studies and is an early example of a symmetry protected topological phase~\cite{HALDANE_PRL_1983}. For the FM case, one can obtain gapped or gapless excitations depending on the sign and size of $ D $ relative to $ J $~\cite{Papaniculou_1987, Papaniculou_1997}. Little is understood about the FM case with $D$$<$$J$, unlike its AFM counterpart. There have been few theoretical studies (e.g.~\cite{Papaniculou_1987, SPIRIN_2003}), and even fewer experiments for this case.

Here we use time-domain THz spectroscopy (TDTS) to experimentally investigate the excitations of NiNb$_{2}$O$_{6}$ and their interactions.  At low temperatures, we find spin excitations whose energies and magnetic field dependence correspond well to the single-magnon spectrum (at $q \sim 0$) of a 1D spin-1 Heisenberg ferromagnetic chain with weak unidirectional anisotropy. At higher temperatures, we observe a renormalization of the magnon energies that depends on the external field direction. This renormalization occurs due to magnon-magnon interactions which are a consequence of the spin-1 nature of the system and do not have an analog in the spin-1/2 chain. To address this, we employ the 
finite temperature dynamical Lanczos algorithm~\cite{Prelovsek2013}, and determine the effect of these interactions on the 
dynamical response at finite temperature. Our findings shed light on the unique nature of magnon interactions for a spin-1 chain and give a general 
perspective on how TDTS in conjunction with numerical calculations can be used to understand finite temperature spin dynamics and interactions.

\begin{figure}
	\centering
	\includegraphics[width=\linewidth]{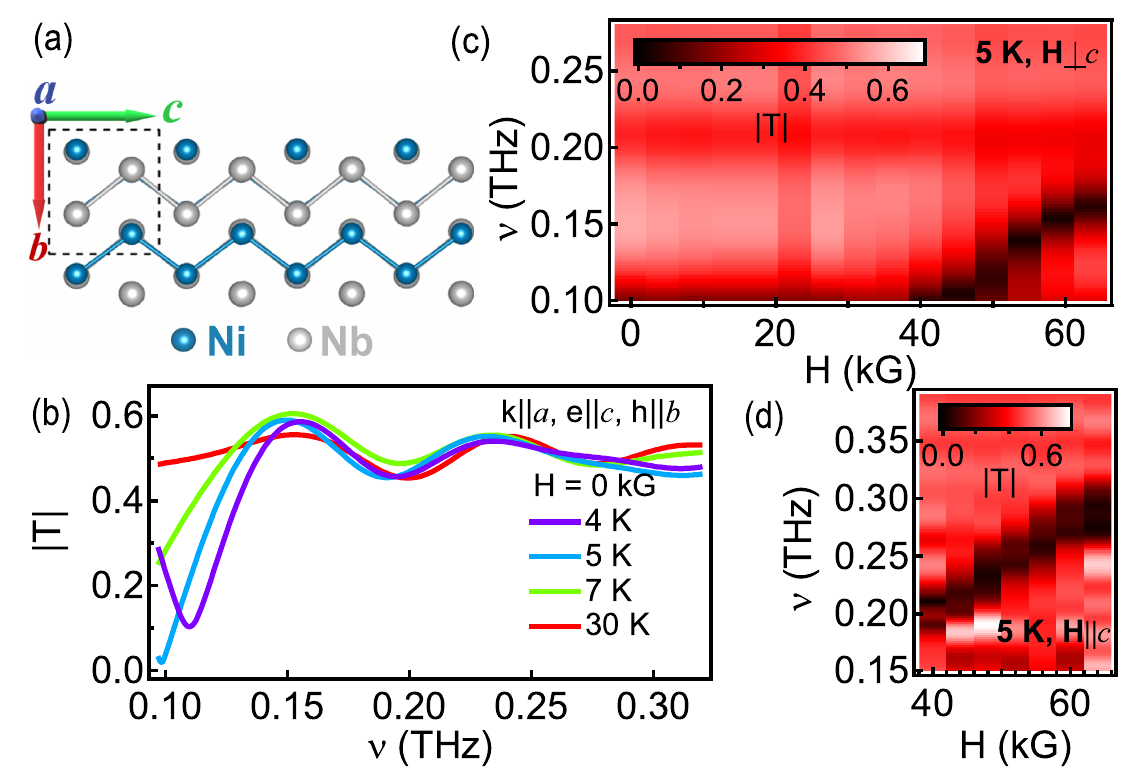}
	\caption{(a) Ni spin-1 chains along the crystallographic $ c $ axis in the $ bc $ plane of NiNb$_{2}$O$_{6}$. (b) Transmission amplitude as a function of frequency $\nu$ in the absence of an external field (H = 0) for various temperatures. Here \textbf{k}$ \parallel $$ a $, \textbf{e}$\parallel$$c$, \textbf{h}$\parallel$$b$, where \textbf{k} is the wave-vector of the incident THz while \textbf{e} and \textbf{h} denote its a.c. electric and magnetic fields respectively. (c)-(d) Field dependence of transmission  at \SI{5}{K} for both transverse and longitudinal field geometries respectively.}
	\label{fig:Crystal} 
\end{figure}

NiNb$_{2}$O$_{6}$ belongs to a family of quasi-1D compounds, the most prominent of which is the Co variant that is perhaps the best example we have of a quasi-1D spin-1/2 Ising system \cite{Coldea_2010,Morris_PRL2014}.  With Ni, the magnetism is both spin-1 and more isotropic.  The structure consists (Fig. \ref{fig:Crystal}(a)) of zigzag edge-sharing chains of NiO$_{6}$ octahedra along the crystallographic $c$ axis with ferromagnetic exchange interactions between nearest-neighbor spin-1 Ni$^{+2}$ ions. Since the intrachain coupling along the $c$ direction is significantly stronger than the interchain coupling ($ J_{\parallel}/J_{\perp}\sim20 $) along the $a$ or $b$ direction, we can consider the system as an effective 1D spin-1 ferromagnetic chain with the $c$ its easy axis~\cite{Heid_1996}. The spin Hamiltonian of this system in an external field can be described with Heisenberg exchange interactions with onsite anisotropy as follows:
\begin{equation}\label{eq1}
H = -|J| \sum_{\langle i,j\rangle } \vec{S_i} \cdot \vec{S_j}  - |D| \sum_{i} (S_i^{z})^2 - g \textbf{H} \cdot \sum_i \vec{S_i} 
\end{equation}
where $ -|J| $ is the ferromagnetic exchange interaction, $D$ is the local onsite uniaxial anisotropy, $S_i$ are spin-1 operators and $g$ is the coupling strength to the external field $\textbf{H}$ (assuming an isotropic g-tensor). Note that in our calculations, $a$, $b$, $c$ refer to crystal directions, whereas $x$, $y$, and $z$ correspond to spin quantization directions. Although an isolated chain with Ising anisotropy orders only at zero temperature, NiNb$_{2}$O$_{6}$'s FM chains order with AFM order below a temperature of \SI{5.7}{K} due to weak interchain interactions. By fitting the specific heat and magnetization data Heid \textit{et} al. \cite{Heid_1996} determined $ J = $ \SI{14.8}{K} (\SI{0.308}{THz}), $ D = \SI{5.2}{K} $ (\SI{0.11}{THz}) and $ g = 2.3 $.

The NiNb$_{2}$O$_{6}$ crystal was grown by the floating zone method and oriented by back reflection Laue diffraction (see Supplementary Material (SM)). TDTS experiments were performed in external fields up to $\mathbf{H}= \SI{68}{kG} $ in both Faraday geometry with transverse field (wave vector \textbf{k}$\parallel$\textbf{H}, \textbf{H}$\perp$$c$) and Voigt geometry with longitudinal field (\textbf{k}$\perp$\textbf{H}, \textbf{H}$\parallel$$c$) at temperatures ranging from $\SI{4}{K}$ to $\SI{150}{K}$.   The spectral range of our TDTS setup is limited to $ >\SI{ 0.1 }{THz} $ in the Faraday and $ >\SI{ 0.15 }{THz} $ in the Voigt geometries.  For magnetic insulators, TDTS functions as high-field electron spin resonance and allows a determination of the complex ac magnetic susceptibility  $ \tilde{\chi}(\nu)=\chi_{1}(\nu) +i\chi_{2}(\nu)$ at THz frequencies in the zero momentum limit. $ \tilde{\chi}(\nu)$ is obtained after normalizing the transmission at a reference temperature (here $\SI{150}{K} $) above the onset of magnetic correlations (see  e.g.~\cite{Kozuki_2011,Pan_Nat_2014,Laurita_thesis_2017,Xinshu_2018}).

Fig. \ref{fig:Crystal}(b) shows the magnitude of the transmission ($T(\nu)$) of NiNb$_{2}$O$_{6}$ as a function of temperature down to \SI{4}{K} with the THz wave-vector \textbf{k}$\parallel$$a$ and the THz ac magnetic field \textbf{h}$\parallel$$b$. In this orientation, a clear absorption peak is observed as the temperature is lowered. The low T peak center frequency of \SI{0.11}{THz} at \SI{4}{K} is in good agreement with  anisotropy parameter $ D=\SI{0.11}{THz} $~\cite{Heid_1996}. Note that in zero-field, the local anisotropy term in $D$~(Eq. \ref{eq1}) breaks the isotropic symmetry of the Heisenberg term resulting in a gap of magnitude $|D|$ in the magnetic excitation spectrum as we observe~\cite{Papaniculou_1987}. To further understand these magnetic excitations, their dynamics and interactions in NiNb$_{2}$O$_{6}$, we perform TDTS measurements as a function of both magnetic field and temperature in both transverse and longitudinal geometries. 

\begin{figure}
	\centering
	\includegraphics[width=3.2in]{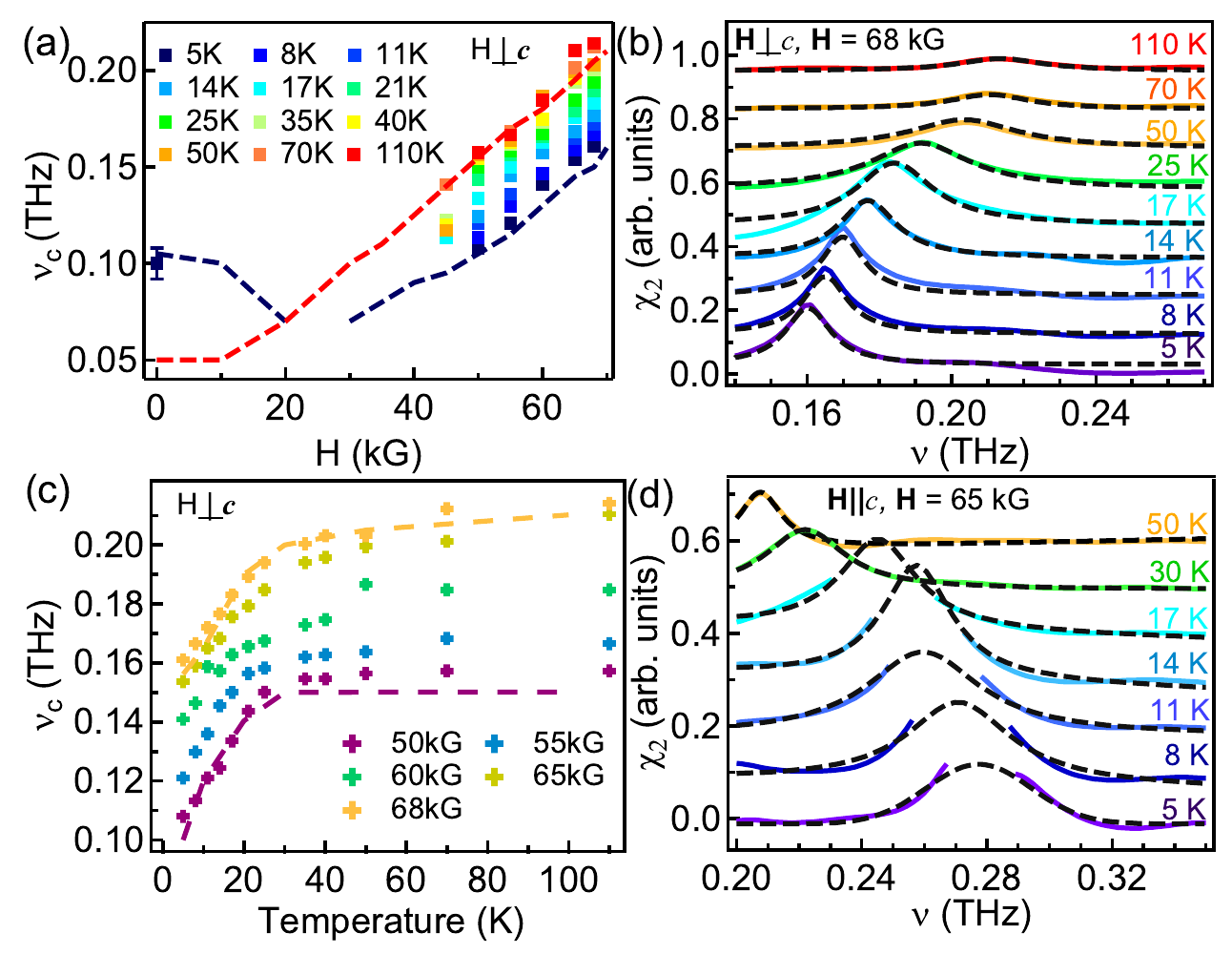}
	\caption{(a) Field and (c) temperature dependence of the center frequency $\nu_c$ of the peak observed in $ \chi_{2} $ for \textbf{k}$ \parallel $$ a $, \textbf{e}$\parallel$$c$, \textbf{h}$\parallel$$ b $. Error bar in (a) represent $ 95{\%}$ confidence interval.  Dashed lines are calculations as described in the text. (b) and (d) Imaginary part of the magnetic susceptibility $ \chi_{2}(\nu) $ for various temperatures. H = 68 kG (H$\parallel$$ a $) in (b) while H = \SI{65}{kG} (H$\parallel$$ c $) in (d). Dashed lines are fit to a Lorentzian. Note that the susceptibility data near the peak is unreliable in (d) due to very strong absorption.   It was excluded from the plots and fits for this reason.}
	\label{fig:faradayexpt} 
\end{figure}

Fig. \ref{fig:Crystal}(c) and Fig. \ref{fig:Crystal}(d) shows the field dependent transmission at \SI{5}{K} in both transverse (\textbf{H}$\perp$$c$) and longitudinal (\textbf{H}$\parallel$$c$) field geometries. Magnon peaks are observed in both cases. The peak center frequency ($\nu_c$) is extracted by fitting the imaginary part of complex magnetic susceptibility, $\chi_{2}(\nu)$, to a Lorentzian (see Fig. \ref{fig:faradayexpt}b and Fig. \ref{fig:faradayexpt}d). $\nu_{c}$ as a function of external magnetic field is shown in Fig. \ref{fig:faradayexpt}(a) for transverse geometry.  At \SI{5}{K} (dark blue squares), only the zero-field spectra and the spectra above \SI{40}{kG} show excitations with $\nu_c > 0.1 $ THz. Above \SI{40}{kG}, $\nu_c$ varies linearly with field as $ \nu_c \sim g\mu_B$ with an offset ($\mu_B $ is the Bohr magneton). From a linear fit to the data at \SI{5}{K} (see SM), we extract a g-factor of $ g=2.14 $ which is in good agreement with Heid \textit{et al.}~\cite{Heid_1996}. The behavior of the magnon center frequency at \SI{5}{K} is consistent with a field-induced ferromagnetic (FM) to paramagnetic (PM) phase transition in the spin-1 chain~\cite{Heid_1996}. To understand the effect of thermal excitations on the magnetic spectra, we measure $\chi_{2}(\nu)$ at fixed field for various temperatures up to \SI{110}{K}.

Fig. \ref{fig:faradayexpt}(b) shows the temperature dependence of $\chi_{2}(\nu)$ at $\SI{68}{kG}$ in the transverse geometry. Clear magnon peaks are observed for temperatures up to $\SI{110}{K}$. At each temperature, the measured magnon $\nu_c$ scales linearly with field for fields $>$ 40 kG (Fig. \ref{fig:faradayexpt}(a)). With increasing temperature, the peak height of the magnon reduces and the peak broadens as expected for thermal broadening of magnetic excitations. Interestingly, rather than staying constant, the magnon peak frequency shifts higher with increasing temperature for each field. In the absence of any transitions, this is quite unusual for magnetic excitations as they typically just broaden with increasing temperature and move only slightly \cite{Morris_PRL2014, Xinshu_2018}. Fig. \ref{fig:faradayexpt}(c) shows the temperature dependence of the magnon $\nu_c$ at different magnetic fields. $\nu_c$ increases linearly with temperature until $\sim\SI{30}{K}$ after which it asymptotically saturates. This agrees well with the temperature above which the susceptibility obeys the Curie-Weiss law (\SI{30}{K}~\cite{Heid_1996}). 

We carry out similar analysis as described above in the longitudinal geometry (\textbf{H}$\parallel$$c$) (see SM). Fig. \ref{fig:faradayexpt}(d) shows the resulting $\chi_{2}(\nu) $ at  $\SI{65}{kG}$ for various temperatures. In this orientation, we also observe magnetic excitations that weaken in intensity with increasing temperature. Importantly, the temperature dependence of the magnon peaks in the longitudinal geometry is opposite to that observed in the transverse geometry.   Here, the magnon peaks shift towards \textit{lower} frequencies with increasing temperature.   Moreover while there is some evidence for a field-induced phase transition in the transverse geometry (See SM), the behavior is different with the longitudinal field (See SM). Magnon center frequencies in both orientations show a linear dependence on field at high fields with a similar g-factors (SM).

\begin{figure}
	\centering
	\includegraphics[width=\linewidth]{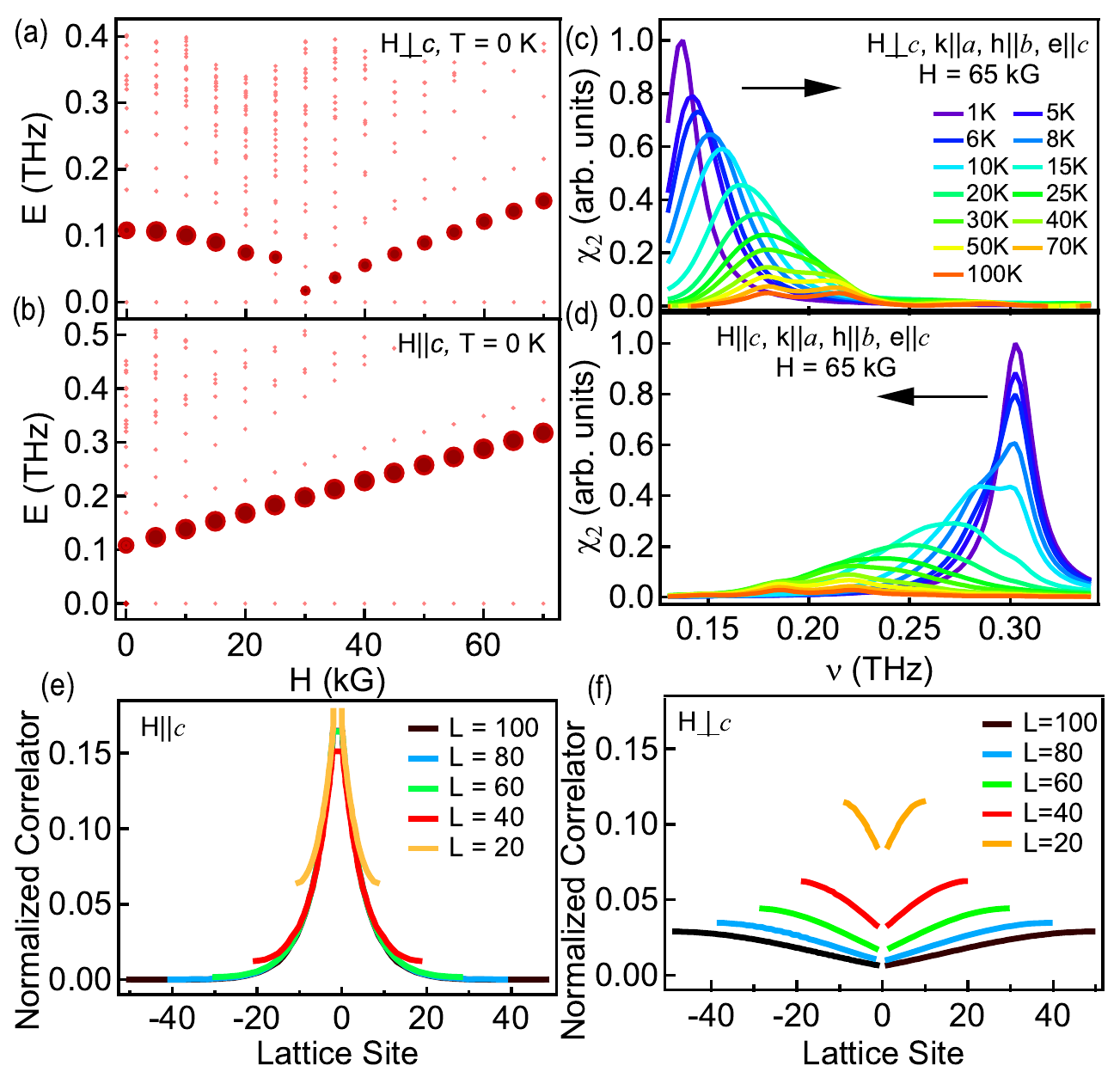}
	\caption{Simulated excited state energy spectra (relative to the ground state) of the spin-1 chain ($L=14$) as a function of external magnetic field (a) along the $ a $-axis and (b) the $ c $-axis. Bold points correspond to the excited states with the size of the big circles proportionate to $|\langle Excited | S^{y} | Ground \rangle|^2$ 
(c)-(d) Simulated temperature dependence of normalized $ \chi_{2} $ computed with finite temperature dynamical Lanczos 
at H = \SI{65}{kG} in both transverse and longitudinal geometries with a Lorentzian used to broaden the spectra.
The normalized magnon-magnon correlator (see text) for (e) transverse (assuming the two spin flip term can be ignored, see SM) and (f) longitudinal geometries with 
respect to a chosen reference site (labeled as site 0), evaluated in the lowest energy 2-magnon wave function in chains of different lengths, 
for $H =$\SI{70}{kG}. The correlator of a site with itself has been omitted. 
}
	\label{fig:faradaysim} 
\end{figure}

We posit that the field-dependent change in the magnon energies at higher temperature is indicative of magnon-magnon interactions renormalizing the excitation spectrum. To check this, we perform exact diagonalization (ED) calculations on the 1D Hamiltonian in Eq. \ref{eq1} for chain length $ L=14 $ at \SI{0}{K} to determine the low-lying energy states of the system as a function of external field. Fig. \ref{fig:faradaysim}(a) and \ref{fig:faradaysim}(b) show these excited state energies with respect to the ground state (GS) for transverse and longitudinal geometries respectively using the parameters $ g=2.14 $, $ J = \SI{14.8}{K} $ and $ D=\SI{5.2}{K} $ determined from earlier heat capacity and magnetization measurements~\cite{Heid_1996}. Because a photon can only excite a magnon with spin change of $\Delta S=1$, at $\SI{0}{K}$ only the first excited states ($E_{10}= E_1 - E_0 $, where $ E_0 $ is the GS energy) are accessible with THz light~\cite{Katsumata_2000}. These states are represented with bold points whose size represents the intensity of the excitations in TDTS experiments. These ED results match closely the measured excitation $\nu_c$ at $\SI{5}{K}$ for both transverse (Fig. \ref{fig:faradayexpt}(a)) and longitudinal geometries (SM). In the transverse case, the ED results suggest a second order transition from the ferromagnetic to a paramagnetic phase, which is analogous to the phase transition in the spin-1/2 transverse field Ising model~\cite{Katsumata_2000}. Since finite size effects near the critical point can be severe, additional DMRG calculations of the first two excited states were performed for a chain length of 200 to confirm this observation (SM). For the longitudinal geometry (Fig. \ref{fig:faradaysim}(b)), the ED calculations show no phase transition, as expected.

Having understood the field dependence of magnons in NiNb$_{2}$O$_{6}$ in the low temperature limit, we now turn to the principal unexpected finding in TDTS measurements, i.e., the unique temperature dependence of magnon energies - shift towards higher (lower) energies with increasing temperature in the transverse (longitudinal) geometries. In the low-temperature limit, it is only possible to excite one magnon due to an absorption of a photon, i.e., by making a transition from the GS to the first excited state ($E_{10}$). However, at higher temperatures ($ T \gtrapprox E_{10} $), it becomes possible to have one-magnon transitions between higher energy states as well. For example, due to thermal excitations of $E_{10}$, a single photon absorption can also excite to two-magnon states with energy $E_{21}= E_2-E_1 $.  In a harmonic model for the magnon spectrum, all excited states should be equally spaced and as such, the excitation peak will always be centered at $\nu_c = E_{10} = E_{21}$ regardless of temperature. However, magnon-magnon interactions can renormalize the excitation spectrum resulting in energy shifts.  To see how this arises in a spin-1 chain, we consider the effects of two-magnon interactions in both field geometries.  For longitudinal field, the energy of one magnon excitation ($ |1\rangle\rightarrow |0 \rangle $) is $ \Delta E_{10}= 2SJ(1-\cos(q))+(2S-1)D + g\mu_B H$~\cite{Papaniculou_1987}, giving $ \Delta E_{10} = D + g\mu_B H$ for $ q=0 $ and $ S=1 $, where the ground state ($\psi_{GS}= |1..111..1\rangle $) has energy $  E_0 = -N(J+D+g\mu_B H)$. At zero-field we get $ E_{10} = D = \SI{0.11}{THz} $ which is exactly the energy of the first excited state observed at \SI{5}{K} [Fig. \ref{fig:Crystal}(b)] in the TDTS measurements and in ED [Fig. \ref{fig:faradaysim}(a)]. 

A two magnon state can be constructed by reducing the azimuthal spin by one unit at two different sites ($ | ..11111.. \rangle \rightarrow |..1011011.. \rangle $) giving an energy of $ 2E_{10} $~\cite{Papaniculou_1987}. This is the case when the two magnons are well separated and not interacting with each other. However, when the two magnons are on adjacent sites,  i.e., $ |00\rangle$, then due to the spin-1 nature of the system the $ |00\rangle$ configuration can tunnel into other spin-preserving states like $ |00\rangle \rightarrow(|$+$1$$-$$1\rangle ~\text{or} ~|$$-$$1$$+$$1\rangle)$ which can lower the total energy of the system. A crude approximation for the energy gained by the $|00\rangle $ magnons due to this hybridization is given by $ E_{gain} = -(J+2D/3)$ which gives $ E_{21} = E_1 - E_0 - 2D/3 < E_{10}$ (See SM). This implies that there is an effective magnon \textit{attractive} interaction creating a two-magnon bound state in longitudinal field. This attraction can also be verified from full diagonalization calculations on long chains up to $L=100$~\cite{ED_note} - we find $E_{20}<2E_{10} $ and from inspecting the spatial magnon-magnon correlator ($L \Big \langle \Big(1-(S^i_z)^2 \Big) \Big(1-(S^0_z)^2 \Big) \Big \rangle$ with respect to the central site chosen to be ``0") for the energetically lowest 2-magnon wavefunction shown in Fig. \ref{fig:faradaysim}(e). Thus, with increasing temperature, we expect a shift of the effective excitation peak to lower frequencies as observed in longitudinal geometry. (Fig. \ref{fig:faradayexpt}(d)). Note that the process described cannot occur in the typically studied spin-1/2 case since two spin-flips (or kinks) cannot tunnel into other spin-preserving configurations.  In this regard spin-1 represents a special situation which is low spin enough to be highly quantum, but yet posses richer internal structure than spin-1/2.

In the transverse geometry when the external field $H$ is large, the GS is non-degenerate and paramagnetic. For transverse field $H$$\gg$$D$, as is the case at $\SI{65}{kG}$, the natural direction for spin quantization is along the $a$ axis. In this case, the energy of the one magnon state is approximately $ E_{10} = -\frac{|D|}{2} + g \mu_B H$ with $ E_0 = -N(J+D/2 + g\mu_BH) $(See SM). This reversal of sign in the $D$ term at large transverse fields, means that  
it costs energy to bring the two magnons adjacent to each other. 
This implies that there is an effective \textit{repulsion} between two magnons in the transverse field paramagnetic phase, this is made more illuminating by observing the spatial magnon-magnon correlator in the energetically lowest 2-magnon wavefunction in Fig. \ref{fig:faradaysim}(f) (SM). Hence with increasing temperature we have 
more repulsive interactions leading to a shift in the effective excitation peak to higher frequencies, which is as observed in the transverse geometry (Fig. \ref{fig:faradayexpt}(c)). 
  
Within the models described above, we can qualitatively explain the observed shift in the magnon energies with temperature in terms of a renormalization of the spectrum based on effective magnon-magnon interactions. To further understand the observations we calculate the finite temperature susceptibility using the 
finite temperature dynamical Lanczos algorithm~\cite{Prelovsek2013} for a chain length $L=14$. For $ q=0 $, we calculate the 
frequency dependent correlation function as $ C^{yy}(\nu,T) = \frac{1}{Z} \sum_{n,m} e^{-\beta E_n} |\langle m|  S^{y} |n \rangle|^2 \delta (E_n + \nu - E_m) $, 
where $ Z$ is the partition function, involving the sum over all eigenenergies, $\beta$ is $ 1/k_BT $, $ E_n$ and $E_m $ are the energies of excited and ground levels respectively. 
At finite temperatures the excited states acquire finite lifetimes due to magnon decay processes.  To compensate for the discrete spectra that arises from finite size effects, we broaden the delta functions using a Lorentzian description $\delta(\nu-\nu_{nm}) = \text{lim}_{\epsilon \rightarrow 0} \frac{\epsilon/\pi}{(\nu-\nu_{nm})^{2} + \epsilon^{2}} $ where $\nu_{nm} = E_m - E_n$ and a broadening $\epsilon=\SI{0.01}{THz}$.   We then calculate the dynamical susceptibility as $ \chi^{yy}(\nu,T)= \pi (1-e^{-\beta \nu}) C^{yy}(\nu,T) $ which is equal to $ \chi_{2}(\nu) $~\cite{Papaniculou_1997}.

Fig. \ref{fig:faradaysim}(c) and \ref{fig:faradaysim}(d) show the simulated $\chi_2(\nu)$ at $\SI{65}{kG}$ at various temperatures for transverse and longitudinal geometries respectively.  In the transverse geometry, the magnon peaks in the calculated $\chi_2(\nu)$ shift towards higher frequencies with increasing temperature while the opposite is the case for the longitudinal geometry (Fig. \ref{fig:faradayexpt}(c) and Fig. \ref{fig:faradayexpt}(d)). For a direct comparison between these calculations and the experiment, we plot the center frequencies of the peaks in the calculated $\chi_2(\nu)$ in the transverse geometry in Figs.\ref{fig:faradayexpt}(a) and (c) (dashed lines) at choice temperatures and fields. There is good agreement between the measured and calculated $\nu_c$. 

To conclude, we have provided experimental and theoretical evidence for magnon interactions in a ferromagnetic spin-1 chain through the observed shift in the peak frequencies with temperature in an external field. Depending on the field orientation these interactions are either attractive or repulsive (at large transverse field). We note that while our experimental work relied on thermal excitations to generate and subsequently probe two magnons within linear response, one can imagine utilizing non-linear THz spectroscopy with intense THz pulses to directly excite higher order states via two-photon absorption \cite{Takayoshi_PRB_1019,wan2019resolving}. Subsequent interaction dynamics may be studied with the resulting non-linear response of the system.

\begin{acknowledgments}
We thank O. Tchernyshyov for helpful conversations.  Work at JHU was supported through the Institute for Quantum Matter, an EFRC funded by the U.S. DOE, Office of BES under DE-SC0019331.  HJC thanks Florida State University and the National High Magnetic Field Laboratory for start up funds and XSEDE resources (DMR190020) and the Maryland Advanced Research 
Computing Center~(MARCC) for computing time. The National High Magnetic Field Laboratory is supported by the National Science Foundation through NSF/DMR-1644779 
and the state of Florida. The DMRG calculations in the SM were performed using the ITensor C++ library (version 2.1.1)~\cite{ITensor}.
\end{acknowledgments}

\clearpage

\setcounter{figure}{0}  
\renewcommand\thefigure{S\arabic{figure}} 
\renewcommand{\figurename}{Fig.} 

\renewcommand{\author}{}
\renewcommand{\title}{Supplementary Material: Probing magnon dynamics and interactions in a spin-1 ferromagnetic chain}
\renewcommand{\affiliation}{}

\onecolumngrid

\begin{center}
	\textbf{\large \title}\\
	\medskip
	\author{Prashant Chauhan$^1$,  Fahad Mahmood$^1$, Hitesh J. Changlani$^{2,3,1}$,  S.~M.~Koohpayeh$ ^1 $, and N. P. Armitage$^1$\\
	\medskip
	\small{$^1$ \textit{The Institute for Quantum Matter, Department of Physics and Astronomy\\The Johns Hopkins University, Baltimore, Maryland 21218, USA}\\
		$^2$ \textit{Department of Physics, Florida State University, Tallahassee, Florida 32306, USA}\\
		$^3$ \textit{National High Magnetic Field Laboratory, Tallahassee, Florida 32304, USA}}}
\end{center}

\twocolumngrid

\section{Sample preparation and crystal structure}
High quality single crystals of NiNb$_{2}$O$_{6}$ (approximately \SI{5}{mm} in diameter) were grown in a four-mirror optical floating zone furnace at Johns Hopkins University. The crystal was cut with its $ a $ axis along the out-of-plane direction of the chain. The sample was polished to a finish of \SI{3}{\micro\meter} and total thickness of $\sim\SI{0.88}{mm}$ using diamond polishing paper and a specialized sample holder to ensure that plane parallel faces were achieved for the THz measurement. 
For the optical experiments the single crystal was oriented using back reflection X-ray Laue diffraction. The sample was mounted on a 4 mm diameter aperture for TDTS measurements.

The spin-1 Heisenberg ferromagnetic chain system NiNb$_{2}$O$_{6}$ crystallizes in the orthorhombic structure with space group \textit{Pbcn}~\cite{Weitzel_1976} 
and lattice constants $ a =\SI{14.038}{\AA} $, $ b = \SI{5.682}{\AA}$ and $ c = \SI{5.023}{\AA} $ at room temperature. The zigzag chain consists of edge-sharing chains 
of NiO$_{6}$ octahedra running along the $c$ axis with ferromagnetic exchange interactions between nearest-neighbor spin-1 Ni$^{+2}$ ions. Each NiO$_{6}$ chain 
is separated by a Nb-O edge-sharing chain in the $bc$ plane~\cite{Heid_1996,Yaegar_PRB1977}. Based on magnetization measurements, the spin easy axis of the Ni$^{+2} $ ions 
is either very close to or coincides with the crystallographic \textit{c} axis~\cite{Yaegar_PRB1977}. There is some disagreement in the canting angle of the magnetic moment relative to $ c $-axis between past studies~\cite{Heid_1996, Yaegar_PRB1977}, but for our purposes we have considered the magnetic moments to be approximately along the $ c $-axis which is also consistent with our observations.

\section{Field and temperature dependence of the excitation energy}
\begin{figure}[ht]
	\includegraphics[width=\linewidth]{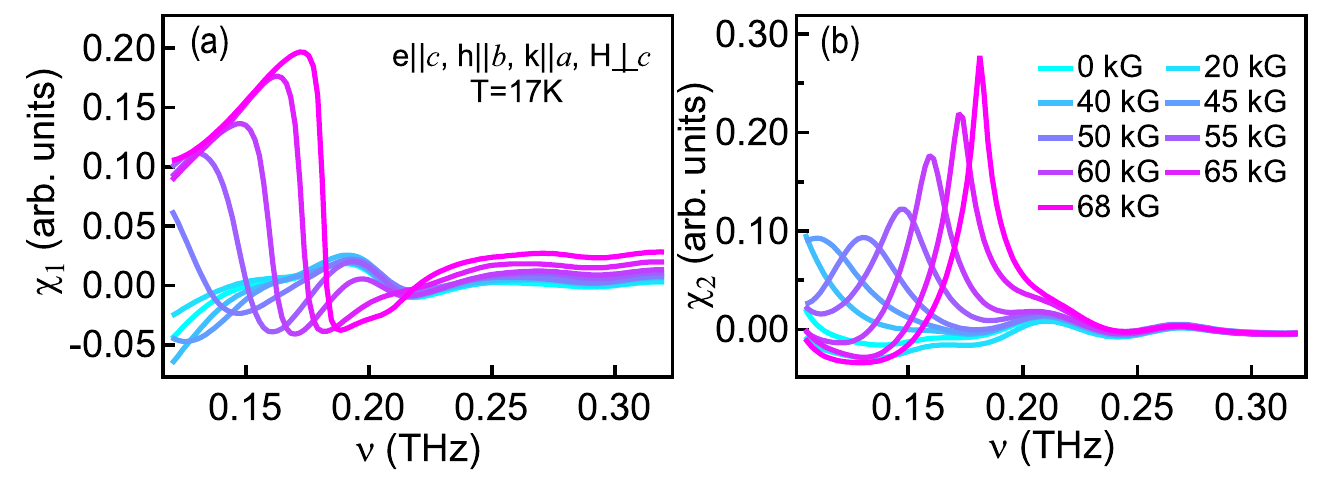}
	\caption{\label{fig:sfig1} Field dependence of (a) real $ \chi_1(\nu) $ and (b) imaginary $ \chi_2(\nu) $ part of magnetic susceptibility spectra at \SI{17}{K} in Faraday geometry when \textbf{H}$\perp$$c$, \textbf{e}$\parallel$$ c$, \textbf{h}$\parallel$$b$, \textbf{k}$\perp$$c$.}
	\vspace{4mm}
\end{figure}

\begin{figure}
	\includegraphics[width=\linewidth]{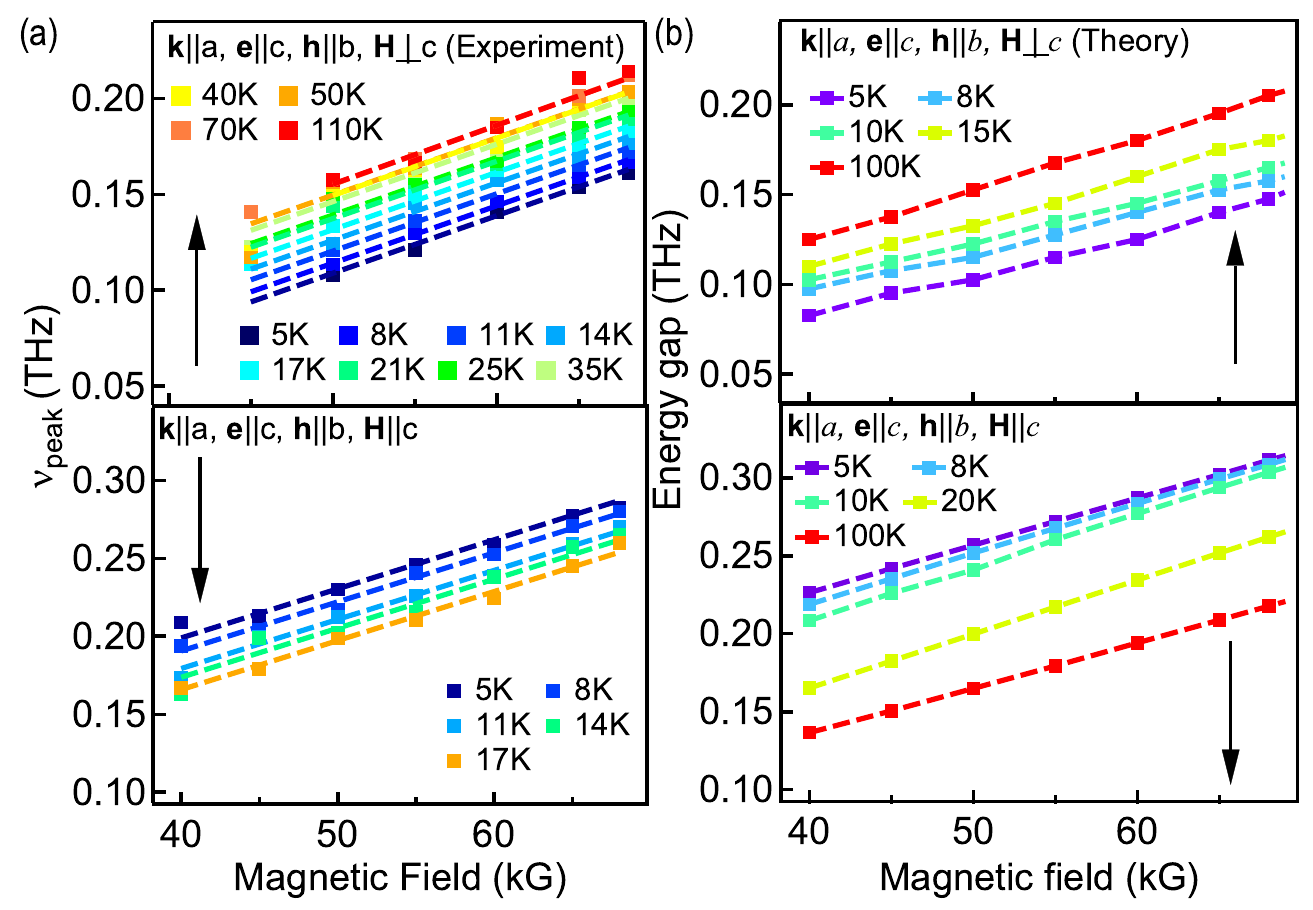}
	\caption{\label{fig:sfig3} Field dependence of the excitation energies for \textbf{h}$\parallel$$b$, \textbf{k}$\parallel$$a$ via peak energies extracted from the $ \chi_2(\nu) $ spectra. (a) Top: peak frequencies for \textbf{H}$\perp$$c$,  Bottom: peak frequencies for \textbf{H}$\parallel$$c$. The dashed lines show fits of peak-frequency vs field. (b) Simulated field dependence of the excitation energy at finite temperature for \textbf{h}$\parallel$$b$, \textbf{k}$\parallel$$c$, when Top: \textbf{H}$\perp$$c$, Bottom: \textbf{H}$\parallel$$c$ for field ranging from 0-70 kG. Arrows indicate the direction in which the peak moves with temperature.}
\end{figure}

Fig.~\ref{fig:sfig1}(a) and (b) show the field dependence of the real and imaginary parts of  $ \tilde{\chi}(\nu) $ respectively,  at \SI{17}{K} in the Faraday geometry. The center frequency, $ \nu_c $, of $ \chi_2(\nu) $ shifts linearly with field above \SI{40}{kG}. Comparing tails of peaks at 0 and \SI{20}{kG} suggests that $ \nu_c(\SI{0}{kG}) > \nu_c(\SI{20}{kG}) $. The absorption peak energies are extracted by fitting the spectra with a Lorentzian. Fig.~\ref{fig:sfig3} shows comparison of experimental and simulated field dependence of the excitation energy at different temperatures in both transverse and longitudinal geometry. The slopes of the linear fits give g-factors of $ g=2.14\pm0.10 $ in the transverse geometry  and $ g=2.25\pm0.10 $ in the longitudinal geometry. The excitation energy increases with temperature in the transverse field case whereas it decreases in temperature in longitudinal field making the temperature dependence of the excitation energy opposite in the two geometries. Dynamical simulations of the excitation energy with field and temperature [Fig.~\ref{fig:sfig3}(b)] gives the same features in high fields (H $>$\SI{40}{kG}) as the experiments in both geometries.
\begin{figure}
	\includegraphics[width=6.5cm]{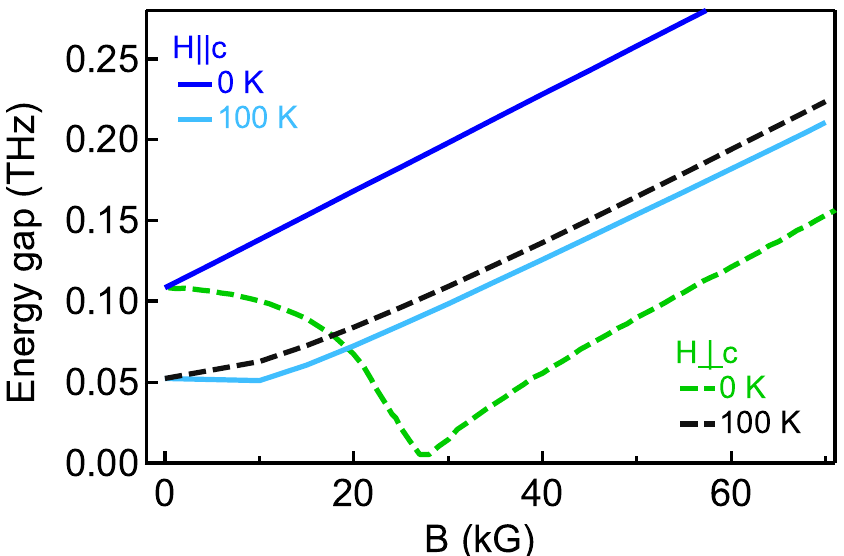}
	\caption{\label{fig:sfig4} Field dependence of excitation energy in transverse \textbf{H}$\perp$$c$  and longitudinal \textbf{H}$\parallel$$c$ geometries.  The \SI{0}{K} data is from DMRG calculations for chain length of $ L=200 $ and the \SI{100}{K} data is from exact diagonalization calculations for $ L=8$.}
\end{figure}

Fig.~\ref{fig:sfig4} shows excitation energies calculated (details below) at two extreme temperatures \SI{0}{K} and \SI{100}{K} cases for both field geometries. For transverse fields (i.e. \textbf{H}$\perp$$c$) the \SI{0}{K} simulation calculated using DMRG shows the field induced phase transition from ferromagnetic to paramagnetic state at \SI{27.9}{kG}, whereas no such phase-transition is seen in the longitudinal field case (i.e. \textbf{H}$\parallel$$c$) and the excitation energy increases linearly with field over the entire range. The \SI{0}{K} data from DMRG calculation was used to get an estimate of the critical point. At high temperatures, of the order \SI{100}{K}, the system is expected to be completely paramagnetic and there is no dependence on direction of the external magnetic field as observed. This comes out naturally from the ED simulations and the excitation energies from both geometries match each other. 

\section{Longitudinal field data fitting}
In the longitudinal (Voigt) geometry our spectrometer has very poor signal-to-noise ratio below \SI{0.3}{THz} and so we only consider the data with transmission above 0.1. The transmission uncertainty below \SI{0.3}{THz} is $ \pm0.1 $ and transmission around the peak value, $ \nu_{c} $, is less than 0.1. Thus we are not able to capture the ($ \nu_{c} $) directly in the experimental susceptibility $ \chi(\nu) $ data. Magnetic susceptibility is calculated as $ \tilde{\chi}(\nu) \approx\frac{2ic}{\sqrt{\epsilon}\nu d}ln(\frac{T_{ref}(\nu)}{T(\nu)})$, where $ T_{ref} $ is transmission above the magnetic ordering temperature and $ d $ is the sample thickness~\cite{Laurita_thesis_2017}. We find $ \nu_c $ by fitting the $ \chi_2(\nu) $ data to a Lorentzian profile [see main text Fig.2].

\section{Temperature dependent dynamics calculations}
We show detailed calculations of $\chi(\nu)$ for temperature dependent dynamics of 1D spin-1 Heisenberg chain in 
transverse as well as longitudinal magnetic fields. In the experiments it has been observed that when the magnetic field is transverse to the chain 
(i.e. \textbf{H}$ \perp$$c$), the peak of $\chi(\nu)$ is found to increase with temperature until it saturates whereas when the field is along chain or 
\textbf{H}$\parallel$$c$ the effective peak frequency is found to reduce till it saturates. Note that in our calculations we use $a$, $b$, $c$ to refer to crystal axes, while
$x$, $y$, $z$ refer to the spin quantization axes. For zero canting angle of the spins (as we have assumed) $c$ points along $z$. 
Defining $S^y = \frac{1}{\sqrt{L}} \sum_{i} S_i^{y}$, we first obtain
\begin{eqnarray}
C^{yy}(t) &=& \langle S^y(t) S^y(0) \rangle \nonumber \\
&=& \text{Tr} (e^{-\beta H} S^y(t) S^y(0))/ \text{Tr} (e^{-\beta H}) \nonumber \\
&=& \frac{1}{Z} \sum_{n} \langle n|  e^{ (-\beta + i t) H} S^{y} e ^{-itH} S^{y} |n \rangle \nonumber \\ 
&=& \frac{1}{Z} \sum_{n,m} e^{-\beta E_n} e^{it (E_n-E_m)} |\langle m|  S^{y} |n \rangle|^2 \label{eq:cyy} 
\end{eqnarray}
where $\beta$ is $ 1/k_B\text{T} $, $Z =\sum_{n} e^{-\beta E_n}  $, $ H $ is the many-body Hamiltonian, $\text{Tr}$ refers to the trace 
and we have introduced a complete set of eigenstates $|n\rangle$,$|m\rangle$ with energies $E_m$, $E_n$ to work with the spectral representation. 

Taking the Fourier transform of Eq.~\eqref{eq:cyy} we get,
\begin{equation}
C^{yy}(\nu) = \frac{1}{Z} \sum_{n,m} e^{-\beta E_n} |\langle m|  S^{y} |n \rangle|^2 \delta (E_n + \nu - E_m)
\end{equation} 
Since $E_n$ and $E_m$ are discrete in any finite quantum system, we broaden the delta functions using a Lorentzian,
\begin{equation}
\delta(\nu-\nu_{nm}) = \text{lim}_{\epsilon \rightarrow 0} \frac{\epsilon/\pi}{(\nu-\nu_{nm})^{2} + \epsilon^{2}} 
\end{equation}
where $\nu_{nm} = E_m - E_n$. Good agreement with experiments is obtained for a broadening factor of $\epsilon=0.01$ THz.

Finally, in order to compare~\cite{Prelovsek2013,Papaniculou_1997}  to $ \chi_2(\nu) $ (in experiments) 
we compute $\chi^{yy}$ as,
\begin{equation}
\chi^{yy}(\nu)= \pi (1-e^{-\beta \nu}) C^{yy}(\nu) 
\end{equation}

For $L=8$, we compute all eigenergies and eigenstates and use the formulae discussed above. For $L=14$, shown in the main text, 
we performed finite temperature dynamical Lanczos calculations, details for which have been extensively discussed in Ref.~\onlinecite{Prelovsek2013}. 
We used 500 Krylov vectors, found to be large enough to supress oscillations in $\chi^{yy}(\nu)$ at high temperature. 
200 random start vectors were used in the averaging procedure involved in this algorithm.

\section{Effective magnon-magnon interactions in the spin-1 chain for magnetic fields applied in the $x$ and $z$ direction}
In the main text we developed a picture for how the magnons attract or repel each other depending on the direction of the applied static magnetic field. 
Here we provide justification for this picture by calculating the approximate change in energy of the excitation due to the 
interaction between two magnons in both longitudinal (\textbf{H}$\parallel$$c$) as well as transverse (\textbf{H}$\perp$$c$) field.
(Note that strictly speaking, a magnon, in the way we have used it here and elsewhere, is well defined only when the total $S_z$ is a good quantum number. 
While this is true for the longitudinal field, it is not so for the transverse field. Yet, in the case of the applied field being much larger 
than $|D|$ this identification can still be made, as will be clarified in the subsequent subsections.)

\subsection{Longitudinal field: Z direction}
Let us consider the case of a single magnon, a single $|0\rangle$ hopping in a ferromagnetic background, for example, 
$ |11..111101111..11\rangle$. When the applied static magnetic field is along the $z$ axis, one can just choose $|1\rangle$,$|0\rangle$,
$|-1\rangle$ to be the usual $S_z$ quantum numbers because the quantization and $\textbf{H}$-field axis coincide. 
The Heisenberg term in this basis is, 
\begin{equation}
-|J| \sum_{\langle i,j\rangle } \vec{S_i} \cdot \vec{S_j} = - \frac{|J|}{2} \Big( S_i^{+} S_j^{-} + S_i^{-}S_j^{+} \Big) - |J| S_i^{z} S_j^{z}.
\end{equation}
The ground state is two fold degenerate and only the $ S_i^{z}S_j^{z} $ term contributes to the energy, in addition to the contributions from $D$ and $H_z$, yielding,
\begin{equation}
E_{GS}=E_0 = -L|J| - L|D| - L g \mu_B H_z
\end{equation} 


Consider now the one magnon excited state (i.e. one $|0\rangle$ in the polarized ground state); in this case
the problem can be mapped to a single ``particle" hopping on the lattice. The exact $k=0$ wave function is, 
\begin{equation}
|\psi\rangle_1 = \frac{1}{\sqrt{L}} \Big( | 0 1 1 1 ....1 \rangle + | 1 0 1 1 .....1 \rangle + | 1 1 0 1 ......1 \rangle ... \Big) 
\end{equation}
The kinetic energy is thus that of a solution to a tight binding model in 1D  $= - 2 |J| \cos(k)$ which equals $-2|J|$ 
for the $k=0$ wave function. The interaction term $S^{i}_zS^{j}_z$ now has two domain walls due to the $...1 0 1...$ pattern, thus this energy is $-(L-2)|J|$. 
The $D$ term contributes $-(L-1)|D|$ and the magnetic field contributes $- (L-1) g \mu_B H_z$. 
Thus, the energy of the $k=0$ 1-magnon state is, 
\begin{align}
E_{1} &= -(L-2)|J| - (L-1) |D| - 2|J| - (L-1) g \mu_B H_z  \nonumber \\
&= -L|J| - (L-1) |D| - (L-1) g \mu_B H_z .
\end{align}
Thus, the energy gap, corresponding to the energy difference of the lowest magnon and the ground state is 
\begin{equation}
E_{10}=E_1 - E_{0} = |D| + g \mu_B H_z.
\end{equation}
In zero field, the gap is thus $|D|=\SI{5.2}{K} = \SI{0.10}{THz}$ which is consistent with that observed in exact diagonalization and the experiment. 
Note that had the Hamiltonian been Ising instead of Heisenberg, we would not have seen the cancellation of the $2|J|$ term and thus the 
lowest excitation would have been $E_{10}=2|J|+|D|$. This would have been detected in our experiments.

Let us now consider the case of the two-magnon excitation. If the two magnons (``0"s) are far apart and never interact, their energy with respect to the 
ground state is just that of two independent magnons i.e. $ 2 |D| + 2 g \mu_B H_z$. However, when two magnons approach each other and are on 
adjacent sites (yielding $ |00\rangle$) then, due to the spin-1 nature of the system, tunneling processes 
$ |00\rangle \rightarrow(|1$$-$$1\rangle ~\text{or}~ |$$-$$11\rangle)$ can lower the energy of the system. 

We estimate the energy that the $ |00\rangle$ magnons gain by hybridizing with each other. 
We set up a local Hamiltonian of two sites in the $|0 0 \rangle$, $|1 $$-$$1\rangle$,$|$$-$$1 1\rangle$ basis
\begin{equation}
\bf{H} \equiv
\left(\begin{array}{ccc}
0 & - J & -J \\
-J & +J - 2D & 0 \\
-J &  0 & +J-2D \end{array} \right) 
\end{equation}
where $J$ and $D$ are positive. Note that the magnetic field contributions are absent. 
We find the eigenvalues $\lambda$ of this matrix, which are solutions to the equation 
\begin{eqnarray}
-\lambda(J-2D-\lambda)^2 - 2J (J-2D-\lambda) = 0 
\end{eqnarray}
are,
\begin{eqnarray}
\lambda &=& J - 2D \nonumber \\
\lambda &=& \frac{1}{2} \Big( J-2D \pm \sqrt{9J^2 + 4D^2 - 4 J D}\Big)
\end{eqnarray}
For the parameter set that is relevant for the material $J-2D$ is positive. 
If $ J\gg D $ we can simplify the other two solutions (ignoring $D^2/J$ terms),
\begin{eqnarray}
\lambda_+ &\approx& 2 J  - \frac{4}{3} D \nonumber \\
\lambda_- &\approx& - J  - \frac{2}{3} D
\end{eqnarray}
The lowest eigenvalue, in this approximation, is thus, $\lambda_- \approx -J - \frac{2}{3} D$. 

\begin{figure}
	\includegraphics[width=\linewidth]{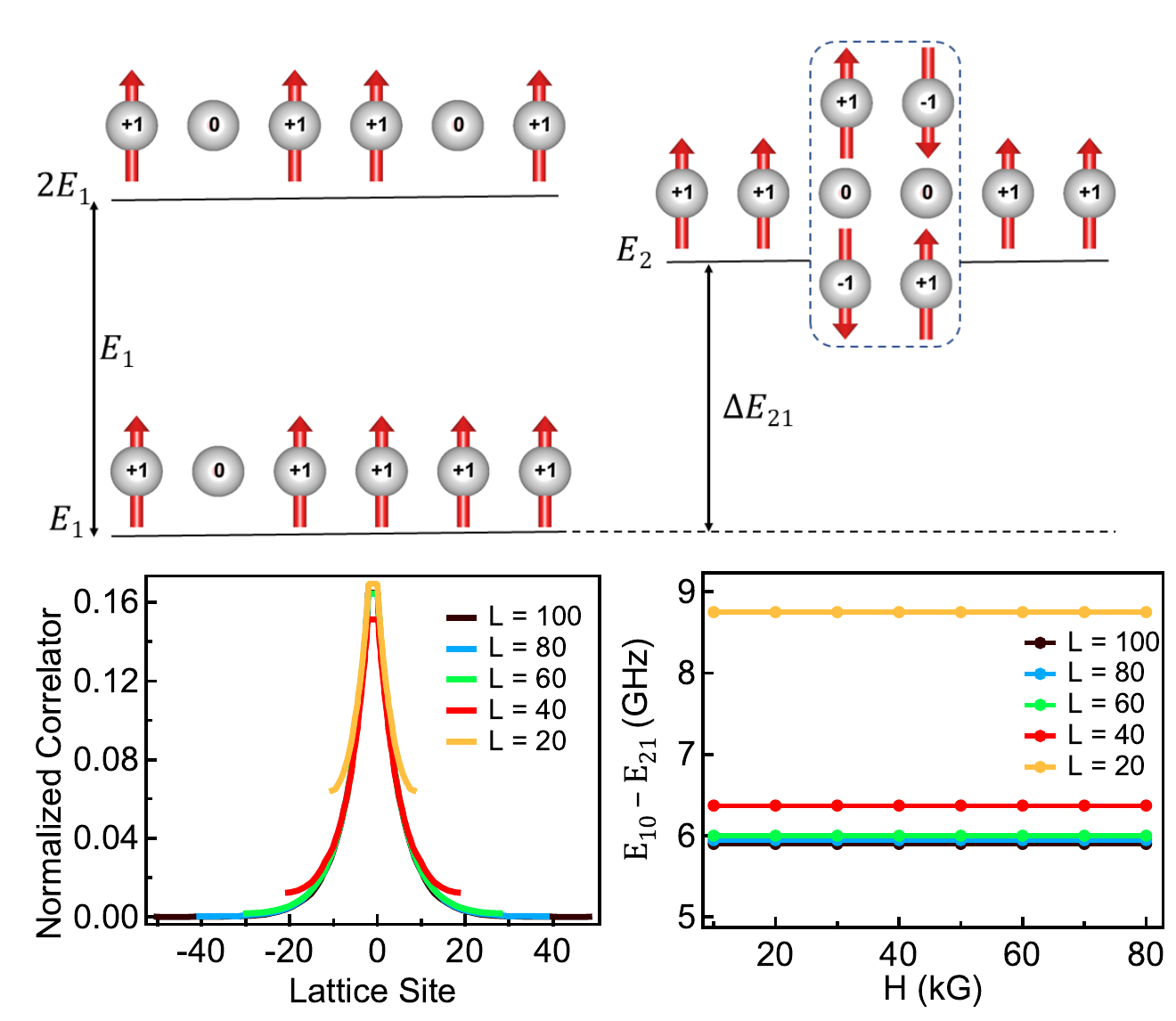}
	\caption{\label{fig:fits} (a) Schematics showing hybridization and renormalization of energy level $ E_2 $ in the case of a longitudinal static magnetic field. (b) 
		The normalized magnon-magnon correlator with respect to a chosen reference site (labeled as site 0), evaluated in the lowest energy 2-magnon wave function in chains of different lengths. The correlator of a site with itself has been omitted. The value of $H$ was chosen to be \SI{70}{kG}. (c) 
		The strength of the magnon-magnon interactions (the difference between the 1- to 2-magnon energy and the ground state to 1-magnon energy i.e. $E_{21}-E_{10}$). 
		This effective attractive interaction is found to be independent of field strength, consistent with the theoretical expectation.} 
\end{figure}

Now consider the total energy of the lowest 2-magnon state. If the two magnons are bound to each other, they move collectively, and 
their kinetic energy in the lowest $k=0$ state would be $-2J$ instead of $-4J$. Also the Ising (diagonal) contribution for the 
``...11111001111" state is now $-(L-3)J$. The onsite anisotropy contribution is $-(L-2)D$. Thus we have, 
\begin{align}
E_2 &= \lambda_- - 2J - (L-3) J - (L-2) D - (L-2) g \mu_B H_z \nonumber \\
&= - \frac{2}{3}D - L J - (L-2) D - (L-2) g \mu_B H_z 
\end{align}
and thus, the energy difference between the lowest 2-magnon state and the lowest 1-magnon state is 
\begin{equation}
E_{21} = |D| + g \mu_B B_z - \frac{2}{3}D = E_1 - E_0 - \frac{2}{3}D < E_{10}
\end{equation}
which means there is an effective magnon attraction. 

The above argument is simplistic, since we have considered a pair of bound magnons on two sites and moved them as a unit to 
account for their collective kinetic energy. In reality, the two magnons have some spread and are not confined solely to two sites. 
To show that the above arguments capture the essence of the physics i.e. the interactions are indeed attractive, 
and to obtain a length scale associated with it, we perform a full diagonalization calculation in each magnon sector individually. 
Since we are interested in the case of a maximum of two magnons, and since $S_z$ is a good quantum number, 
we have directly diagonalized the two-magnon Hilbert space of periodic chains of length $L=20$ to $L=100$ sites with dimension $L(L-1)/2 + L$. 
(This was done by enumerating all states with two 0s and $(L-2)$ 1s, and one -1 and $(L-1)$ 1s, and constructing the Hamiltonian matrix explicitly in this basis). 

For the lowest energy 2-magnon wavefunction, we plot the normalized magnon-magnon correlator relative to the central site ($c$), 
$L \langle (1-(S^i_z)^2) (1-(S^c_z)^2) \rangle$, as shown in the bottom left panel of Fig.~\ref{fig:fits}. To a very good approximation 
it appears to be independent of $L$, (except for sites that are separated on the scale of $\sim L/2$ itself).
However, the strength of the attraction is much weaker than $\frac{2}{3} D$; this is expected 
because the two magnons are bound over a spacing of the order of 20 sites rather than two sites.   
Importantly, the attraction strength is \textit{independent} of \textbf{H} in this geometry 
as we see in the bottom right panel of Fig.~\ref{fig:fits}.

\subsection{Transverse field: x direction}

In a transverse magnetic field, one expects a quantum phase transition between ferromagnetic and paramagnetic states.  
It is indeed observed in the numerical (ED/DMRG) calculations with a hint of it in an experiment.  Note that in Fig. 1(b) the absorption at the lowest frequencies is lower at 20kG than it is at 0 or 40 kG.   This is consistent with an excitation that softens near 20 kG and a quantum phase transition in this range.  If the transverse field $H$ is small, the ground state is two fold degenerate (all spins point either along $+z$ or $-z$), but for large $H$ the spins point along the field direction and the ground state is non degenerate and paramagnetic. In this 
case the $z$ axis is no longer the quantizing axis for spins in the ground state, rather it is determined by diagonalizing 
the onsite $-DS_z^2 - g \mu_B H S_x $ term. For $H\gg D$, the natural spin quantization axis is $x$ i.e. $x$$\perp$$c$. 

Choosing the quantization axis as the $x$ axis (instead of the conventional $z$ axis), 
the operator $(S_i^{z})^2 $ is
\begin{align}
(S_i^{z})^2 &=& \frac{1}{4} (S_i^{+}S_i^{+} + S_i^{-}S_i^{-}) + \frac{1}{2} (S_i^2 - (S_i^{x})^2) 
\end{align}
The operator $(S_i^{+}S_i^{+} + S_i^{-}S_i^{-})$ is a onsite double raising/lowering one and acts only on $|1\rangle$ or $|-1\rangle$. 
The Hamiltonian now reads, 
\begin{equation}
\begin{split}
H = -|J| \sum_{\langle i,j\rangle } \vec{S_i} \cdot \vec{S_j} + \frac{|D|}{2} \sum_{i} (S_i^{x})^2\\
- \frac{|D|}{4} (S_i^{+}S_i^{+} + S_i^{-}S_i^{-}) - g \mu_B H_x \sum_i S_i^x
\end{split}
\end{equation}
where we have dropped the constant term $-\frac{1}{2} |D| S_i^2$  which just leads to an overall constant shift of the 
energy for all eigenstates. 

\begin{figure}[htpb]
	\includegraphics[width=7.00cm]{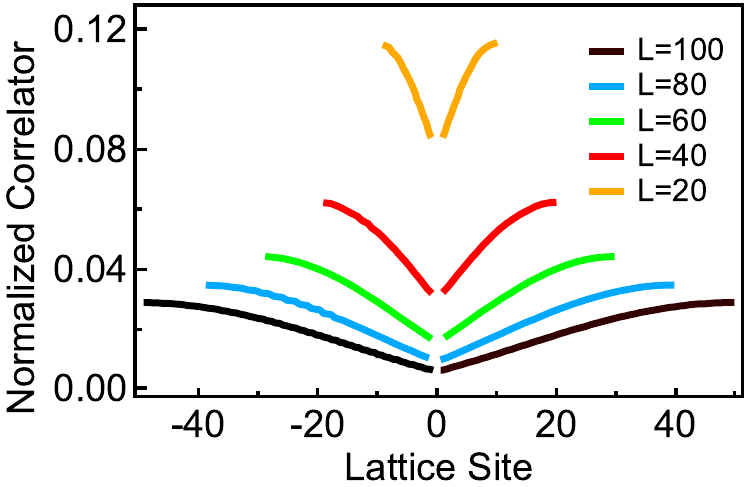}
	\caption{\label{fig:sfig6} Normalized magnon-magnon correlator with respect to a chosen reference site (labelled as site 0), 
		evaluated in the lowest energy 2-magnon wave function in chains of different lengths for the case of a static transverse magnetic field, assuming $H \gg D$. 
		The correlator of a site with itself has been omitted. The value of $H$ was chosen to be \SI{70}{kG}. 
		The dip in the value of the correlator at site 0 and larger value further away from it 
		shows that the second magnon repels the magnon located at site 0.} 
	\label{fig:sfig7}
\end{figure}

The term $(S_i^{+}S_i^{+} + S_i^{-}S_i^{-})$ changes the magnon sector by two, and couples the 1-magnon states to 3 magnon states, 
and the 2 magnon states to the 4 magnon states and 0 magnon states (i.e. the ground state). Since these states are separated by a large energy scale $H$, the 
effect of these states on the energy spectrum is expected to be of the order of $D^2/H$, and thus we drop this term for a qualitative analysis. 

With this term dropped, this Hamiltonian maps to the Hamiltonian of the $z$ case with $-|D| \rightarrow +|D|/2$; 
the magnon attraction is replaced by a magnon repulsion. The lowest 2-magnon state, sees only a weak repulsion of the two 
magnons, since in the thermodynamic limit the magnons can completely avoid each other (their kinetic energy is unaffected by each other). 
This is captured in the normalized correlator shown in Fig.~\ref{fig:sfig7} for various lattice sizes.

Our arguments have considered only the $k=0$ contributions (ground state to lowest 1-magnon and lowest 1-magnon to lowest 2-magnon), 
for both field directions. But at finite temperature, there are many more contributions to $\chi(\nu)$ 
from the 1-magnon to 2-magnon and 2-magnon to 3-magnon transitions that have $k \neq 0$, but still respect the selection rule for matrix elements 
that contribute to $\chi(\nu)$ i.e. the \textit{momentum difference} between states is zero ($\delta k = 0$).
For example, in the $x$ case, even though the 2-magnon state with momentum $k=0$ avoids repulsive interactions efficiently, 
the thermally populated $k=0$ 1-magnon states will couple to their 2-magnon counterparts with the same momentum, here the two 
individual magnons with momenta $k'$ and $k-k'$ will see effective repulsive interactions. Since the phase space of these contributions 
is large, these transitions will dominate at finite temperature on the order of the 1-magnon bandwidth $\sim J$. 
A detailed analysis of the contributions of these processes to the susceptibility will be addressed elsewhere.


\begin{thebibliography}{43}
\bibitem [{ED_()}]{ED_note}%
	\BibitemOpen
	\href@noop {} {}\bibinfo {note} {When only the 2-magnon sector is of
		interest, we have performed exact diagonalizations for chain lengths of up to
		100 sites. See SM for more details.}\BibitemShut {Stop}%
\end{thebibliography}
\end{document}